\documentclass[runningheads,a4paper]{llncs}

\usepackage{amssymb}
\usepackage{graphicx}
\usepackage{balance}
\usepackage{multirow}
\usepackage{hyphenat}
\usepackage{url}
\usepackage{hyperref}

\urldef{\mailsa}\path|{anderson, isabel.suarez, david}@uni-kassel.de|
\urldef{\mailsb}\path| yaqian.xu@comtec.eecs.uni-kassel.de|     
\newcommand{\keywords}[1]{\par\addvspace\baselineskip
\noindent\keywordname\enspace\ignorespaces#1}

\begin{document}

\mainmatter 

\title{An Ontology-Based Reasoning Framework for Context-Aware Applications\thanks{The final publication is available at Springer via~\url{http://dx.doi.org/10.1007/978-3-319-25591-0_34}}}

\titlerunning{Framework for Context-Aware Applications}

\author{Christoph Anderson\inst{1}, Isabel Suarez\inst{1}, Yaqian Xu\inst{2}, \and Klaus David\inst{1}}

\authorrunning{Anderson et al.}

\institute{University of Kassel, Chair of Communication Technology\\
Wilhelmsh\"oher Allee 73, 34121 Kassel, Germany\\
\inst{1}\mailsa\\
\inst{2}\mailsb\\
\url{http://www.comtec.eecs.uni-kassel.de/}}

\maketitle

\begin{abstract}
Context-aware applications process context information to support users in their daily tasks and routines. These applications can adapt their functionalities by aggregating context information through machine-learning and data processing algorithms, supporting users with recommendations or services based on their current needs. In the last years, smartphones have been used in the field of context-awareness due to their embedded sensors and various communication interfaces such as Bluetooth, WiFi, NFC or cellular. However, building context-aware applications for smartphones can be a challenging and time-consuming task. In this paper, we describe an ontology-based reasoning framework to create context-aware applications. The framework is based on an ontology as well as micro-services to aggregate, process and represent context information.
\end{abstract}

\section{Introduction}
In the last few years, mobile phones have evolved from devices, used for voice communication and sending text messages only, to powerful smartphones with multiple embedded sensors and communication interfaces such as WiFi, Bluetooth, NFC or cellular. With their increasing processing capabilities and sensors such as accelerometers, gyroscopes, and magnetometers, smartphones are now being used for internet browsing, social networking, playing games, watching videos or listening to music \cite{Falaki.2010}. Other fields of smartphone applications are activity recognition and context-awareness. In the area of activity recognition, for example, smartphones are used to recognize Activities of Daily Living (ADL) \cite{Roy.2013},\cite{Shoaib.2013}. Context-aware applications can be utilized in the field of Ambient-Assisted-Living (AAL) and home automation. These types of applications monitor the environment to aggregate context information to provide recommendations or services to the user. Typically, this information is extracted from sensor data, represented as time series, by using machine-learning and data processing algorithms. Existing frameworks in this field can be divided into two categories based on the way contexts are represented. The first category includes frameworks that describe context information without semantic e.g. as plain programming objects such as strings or class objects using object-oriented models \cite{Sridevi.2012},\cite{Raento.2005},\cite{Carlson.2012},\cite{Ferreira.2015}. However, these object-oriented models are not suitable for knowledge and data sharing in heterogeneous pervasive environments \cite{Chen.2004}. The second category consists of frameworks supporting semantic representations of context information \cite{GuPZ04},\cite{Meditskos.2013},\cite{Paspallis.2014}. These frameworks exploit ontologies, first-order logics or other description technologies to represent context information semantically. Building context-aware applications on smartphones that represent context information semantically as well as aggregate and process contexts through sensor information is still a challenging task.\\
In this paper, we present an ontology-based framework to create context-aware applications. In addition to the ontology, we extend the current state of the art frameworks by integrating micro-services to aggregate and process context information. By supporting reasoners such as Pellet \cite{Sirin.2007}, HermiT \cite{Glimm.2014} and JFact \cite{Tsarkov.2006}, the framework can deduce complex contexts from already aggregated context information by using the reasoning paradigm of the Web-Ontology-Language (OWL). In the following sections, we present the architecture of our framework including a schematic overview over its components. A discussion about the limitations of the framework and a conclusion with a summary of future work is given at the end of this paper.

\section{Framework-Architecture}\label{section_architecture}
\begin{figure}[ht]
  \centering
  \includegraphics[scale=0.6]{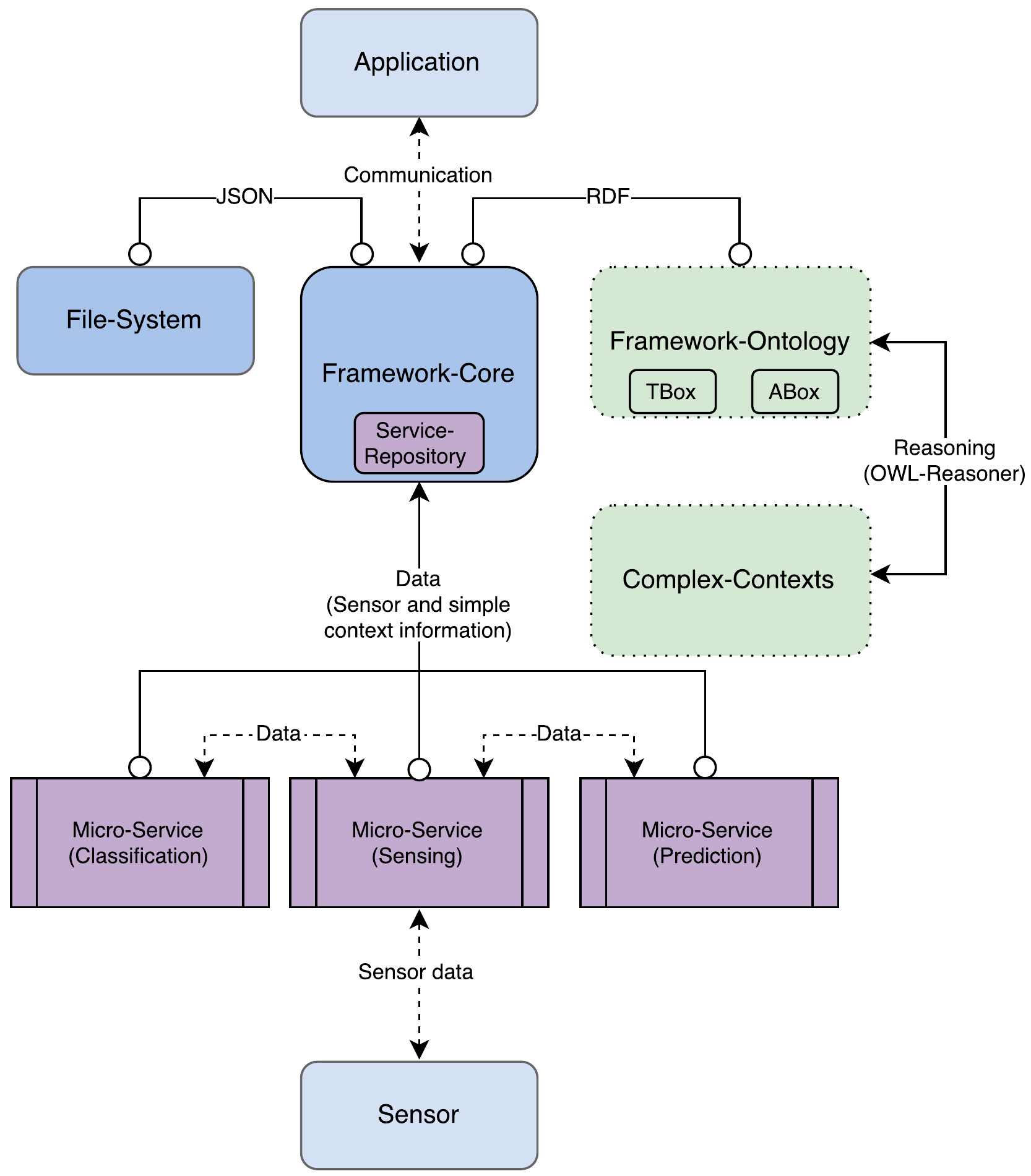}
  \caption{Schematic view of the framework architecture}
  \label{fig:architecture}
\end{figure}

The framework is based on an OWL ontology to model and represent context information. By using an ontology, complex contexts can be deduced by using the reasoning paradigm of OWL. In combination with the ontology, the framework exploits micro-services to aggregate and process context information from embedded smartphone or environmental sensors. Micro-services refer to standalone applications providing a background service only. These background services implement one or multiple functionalities such as context classification, prediction or sensor data aggregation, which can be used by applications or other micro-services. A schematic view of the framework architecture is given in Figure \ref{fig:architecture}.

\subsection{Micro-Services} \label{sub_micro_services}
The micro-service architectural style is a mechanism, where a dedicated application provides one or multiple services, each one of them running in its process. In our framework, services communicate over an Inter-Process-Communication interface (IPC) with other services or the framework core. Figure \ref{fig:architecture} depicts three possible types of micro-services. Since the framework has to exploit sensors to aggregate and process sensor data, the first type of micro-services (\textit{Sensing}) is used to establish a connection to different sensors. Either embedded smartphone sensors such as the accelerometer, gyroscope or magnetometer or environmental sensors such as movement sensors or door contacts can be used. Establishing a connection to environmental sensors, for example, can be done by implementing a micro-service requesting a web-service abstracting sensor information e.g. in a smart home environment. Already aggregated sensor data can be distributed to other micro-services or the framework core by IPC. The second type of micro-services (\textit{Classification}) is used to classify context information from sensor data. Existing approaches such as \cite{Kwapisz.2010} can be implemented as services to classify context information such as standing, sitting, walking, and lying. The third type of micro-services (\textit{Prediction}) is used for context prediction. Generally, existing context prediction algorithms such as \cite{Sigg.2010} or \cite{Gopalratnam.2004} exploit context histories to build prediction models. These algorithms predict future contexts providing further information about the next context of the user. To support context prediction algorithms, the framework provides access to already aggregated context information in form of a context history stored on the smartphone's file-system.

\subsection{Framework core}\label{sub_framework_core}
The framework core is the central component of the framework. It maintains the ontology and provides an interface to establish IPC connections to installed micro-services. The interface controls the life-cycle of micro-services. For this reason, the core maintains a service-repository of installed micro-services in the form of a database. The repository is needed to establish connections to installed services. If the user installs or removes a service, the framework core updates the corresponding entry in the database. Additionally, the framework core process information from micro-services such as aggregated contexts, raw sensor data or context predictions. While most context classification or prediction approaches rely on sensor or context histories to classify or predict new context information, the framework core saves this information on the smartphone's local file-system. Already classified contexts are added to the ontology.

\subsection{Framework ontology}\label{sub_framework_ontology}
The framework uses an OWL ontology to model and represent context information. The ontology describes the terminology of the context model in the form of conceptual classes and relationships between these classes. These classes and relations define the structure of the underlying model and are stored inside a terminological box (TBox). Instances of conceptual classes, their attributes and relationships are stored inside an assertional box (ABox). With the reasoning paradigm of OWL ontologies, the framework can deduce complex contexts from already aggregated information. In a smart environment, for example, the ontology can infer that a user is making coffee by using dense sensing \cite{Chen.2012}.

\section{Limitations}
An experimental evaluation of the framework was carried out by implementing a context-based communication filter for Android based smartphones. The filter blocks certain communication such as phone calls, text messages or emails by aggregating and processing context information of the user in a smart home environment located in our department. Similar to the evaluation in \cite{Yus.2013}, we experienced memory and performance issues on the tested smartphones (Google Nexus 4, 5 and 6) when reasoning larger ontologies\footnote{More than 6.000 Axioms}. Besides memory and performance issues, applications using the framework may have time constraints when deducing new context information from already aggregated contexts. These time constraints may be violated when using reasoners that are not optimized for large ontologies or when using power-limited smartphones. Besides the limitations caused by different ontology sizes and reasoners being used, efforts regarding the privacy and security of context information have to be made. Although, context information is stored on the smartphone's local file-system, the exchange of data e.g. using environmental sensors, has to be secured by using current state of the art encryption technologies such as TLS (Transport Layer Security).

\section{Conclusion}
Context-aware applications process context information to support users in their daily tasks and routines. Building context-aware applications for smartphones that represent context information semantically as well as aggregate and process contexts through smartphone or environmental sensor is still a challenging task. In this paper, we presented an ontology-based framework to create context-aware applications. By utilizing an ontology, context information can be described semantically. In addition to the ontology, we implemented micro-services to aggregate and process context information from embedded smartphone or environmental sensors. By utilizing reasoners such as Pellet, HermiT or JFact, the framework can deduce complex contexts from already aggregated context information by using the reasoning paradigm in OWL. In the future, we plan to publish the framework, making it available to the community. Also, we plan to take into account semantic rule languages such as SWRL (Semantic-Web-Rule-Language), as well as other reasoners to extend the functionalities of the framework.

\section{Acknowledgements}
This work has been co-funded by the Social Link Project within the Loewe Program of Excellence in Research, Hessen, Germany

\balance
\bibliographystyle{splncs03}
\bibliography{data/library}
\end{document}